\documentclass[12pt]{article}
  \usepackage{amsfonts}
  \usepackage{amsmath}
\usepackage{amssymb}
\usepackage{amscd}
\usepackage[dvips]{graphicx}

\begin{document}

~~
\bigskip
\bigskip
\begin{center}
{\Large {\bf{{{Quantum mechanics of many particles defined on
twisted N-enlarged Newton-Hooke space-times}}}}}
\end{center}
\bigskip
\bigskip
\bigskip
\begin{center}
{{\large ${\rm {Marcin\;Daszkiewicz}}$ }}
\end{center}
\bigskip
\begin{center}
{ {{{Institute of Theoretical Physics\\ University of Wroc{\l}aw pl.
Maxa Borna 9, 50-206 Wroc{\l}aw, Poland\\ e-mail:
marcin@ift.uni.wroc.pl}}}}
\end{center}
\bigskip
\bigskip
\bigskip
\bigskip
\bigskip
\bigskip
\bigskip
\bigskip
\begin{abstract}
We provide the quantum mechanics  of many particles moving in
twisted N-enlarged Newton-Hooke space-time. In particular, we consider
the example of such noncommutative system - the set of M
particles moving in Coulomb field of external point-like source and interacting each other also by Coulomb potential.
\end{abstract}
\bigskip
\bigskip
\bigskip
\bigskip
\eject

\section{Introduction}

The idea to use noncommutative coordinates is quite old - it goes
back to Heisenberg and was firstly formalized by Snyder in
\cite{snyder}. Recently, however,  there were  found new formal
arguments based mainly on Quantum Gravity \cite{2}, \cite{2a} and
String Theory models \cite{recent}, \cite{string1}, indicating that
space-time at Planck scale  should be noncommutative, i.e. it should
have a quantum nature. On the other side, the main reason for such
considerations follows from many phenomenological considerations,
which state that relativistic space-time symmetries should be
modified (deformed) at Planck scale, while  the classical Poincare
invariance still remains valid at larger distances
\cite{1a}, \cite{1d}.

It is well-known that  a proper modification of the Poincare and
Galilei Hopf algebras can be realized in the framework of Quantum
Groups \cite{qg1}, \cite{qg3}. Hence, in accordance with the
Hopf-algebraic classification  of all deformations of relativistic
and nonrelativistic symmetries (see \cite{class1}, \cite{class2}),
one can distinguish three
types of quantum spaces \cite{class1}, \cite{class2} (for details see also \cite{nnh}):\\
\\
{ \bf 1)} Canonical ($\theta^{\mu\nu}$-deformed) type of quantum space \cite{oeckl}-\cite{dasz1}
\begin{equation}
[\;{ x}_{\mu},{ x}_{\nu}\;] = i\theta_{\mu\nu}\;, \label{noncomm}
\end{equation}
\\
{ \bf 2)} Lie-algebraic modification of classical space-time \cite{chiqft}-\cite{lie1}
\begin{equation}
[\;{ x}_{\mu},{ x}_{\nu}\;] = i\theta_{\mu\nu}^{\rho}{ x}_{\rho}\;,
\label{noncomm1}
\end{equation}
and\\
\\
{ \bf 3)} Quadratic deformation of Minkowski and Galilei  spaces \cite{chiqft}, \cite{chiqft1}, \cite{lie1}-\cite{paolo}
\begin{equation}
[\;{ x}_{\mu},{ x}_{\nu}\;] = i\theta_{\mu\nu}^{\rho\tau}{
x}_{\rho}{ x}_{\tau}\;, \label{noncomm2}
\end{equation}
with coefficients $\theta_{\mu\nu}$, $\theta_{\mu\nu}^{\rho}$ and  $\theta_{\mu\nu}^{\rho\tau}$ being constants.\\
\\
Besides, it has been demonstrated in \cite{nnh}, that in the case of
so-called N-enlarged Newton-Hooke Hopf algebras
$\,{\mathcal U}^{(N)}_0({ NH}_{\pm})$ the twist deformation
provides the new  space-time noncommutativity of the
form\footnote{$x_0 = ct$.},\footnote{ The discussed space-times have been  defined as the quantum
representation spaces, so-called Hopf modules (see e.g. \cite{oeckl}, \cite{chi}), for quantum N-enlarged
Newton-Hooke Hopf algebras.}
\begin{equation}
{ \bf 4)}\;\;\;\;\;\;\;\;\;[\;t,{ x}_{i}\;] = 0\;\;\;,\;\;\; [\;{ x}_{i},{ x}_{j}\;] = 
if_{\pm}\left(\frac{t}{\tau}\right)\theta_{ij}(x)
\;, \label{nhspace}
\end{equation}
with time-dependent  functions
$$f_+\left(\frac{t}{\tau}\right) =
f\left(\sinh\left(\frac{t}{\tau}\right),\cosh\left(\frac{t}{\tau}\right)\right)\;\;\;,\;\;\;
f_-\left(\frac{t}{\tau}\right) =
f\left(\sin\left(\frac{t}{\tau}\right),\cos\left(\frac{t}{\tau}\right)\right)\;,$$
$\theta_{ij}(x) \sim \theta_{ij} = {\rm const}$ or
$\theta_{ij}(x) \sim \theta_{ij}^{k}x_k$ and  $\tau$ denoting the time scale parameter
 -  the cosmological constant. It should be also noted that different relations  between all mentioned above quantum spaces ({\bf 1)}, { \bf 2)}, { \bf 3)}
and { \bf 4)}) have been summarized in paper \cite{nnh}.

Recently, there appeared a lot of papers dealing with classical
(\cite{deri}-\cite{daszwal}) and quantum (\cite{qm1}-\cite{oscy})
mechanics, Doubly Special Relativity frameworks (\cite{dsr1a},
\cite{dsr1b}), statistical physics (\cite{maggiore}, \cite{rama})
and field theoretical models (see e.g. \cite{przeglad}), defined on
quantum
 space-times (\ref{noncomm}),
 (\ref{noncomm1})\footnote{For earlier studies see \cite{lukiluk1} and
\cite{lukiluk2}.}. Particulary, there was
 investigated the impact of the mentioned
above   deformations  on dynamics of basic classical and quantum
systems. Consequently, in  papers \cite{romero}, \cite{romero1}, the
authors considered  classical particle moving in central
gravitational field defined on canonically deformed space-time
(\ref{noncomm}). They  demonstrated, that in such a case there
is generated Coriolis force acting additionally on the moving
particle. Besides, in articles \cite{lodzianieosc}, \cite{romero}
and \cite{oscy} there was analyzed classical and quantum oscillator
model formulated on canonically and Lie-algebraically deformed
space-time respectively. Particulary, there has been found its
deformed energy spectrum as well as the corresponding equation of
motion. Interesting results have been also obtained in two papers
\cite{qm1} and \cite{qm2} concerning the  hydrogen atom model defined
on quantum space (\ref{noncomm}). Besides,
it should be noted that  there appeared article \cite{toporzelek},
which provides the link between Pioneer anomaly phenomena
\cite{piophen} and classical mechanics defined on $\kappa$-Galilei
quantum space. Precisely,  there has been demonstrated that
additional force term acting on moving satellite can be identified
with the force generated  by space-time noncommutativity. The value
of deformation parameter $\kappa$ has been fixed  by comparison of
obtained theoretical results with observational data.

Unfortunately, in all mentioned above articles there were analyzed
only the one-particle classical and quantum dynamics in
the field of forces. Here, we extend  such a kind of investigations
to the quantum mechanics of many particles, which move in the
modified twist-deformed N-enlarged Newton-Hooke space-time
\begin{equation}
[\,t,x_{iA}\,] = 0\;\;\;,\;\;\;[\,x_{iA},x_{jB}\,] =
if(t) = if_{\pm}\left(\frac{t}{\tau}\right)\theta_{ij}\;\;\;;\;\;\;i, j = 1,2, 3\;, \label{canamm}
\end{equation}
with  indices $A, B = 1,2, \ldots ,M$ labeling the particle.
Further, we indicate that as in the case of one-particle quantum system
there appeared additional dynamical terms generated by space-time
noncommutativity. Of course,  in the case of Coulomb potential for $M=1$ and $f(t) = \theta_{ij}$ our results become the same
as the ones obtained in \cite{qm1} and \cite{qm2}  respectively.

The motivations for present studies are manyfold. First of all we
extend in natural way the  results for quantum one-particle model to the much more complicated
many-particle system. Secondly, such investigations permit to
analyze the deformations of wide class of physical models such as, for
example, the noncommutative many-electron atoms or noncommutative many-atomic molecules \cite{24}, \cite{24a}. Finally, it gives a starting
point for the construction of Dirac quantum mechanics for
many particles defined on the relativistic counterpart of modified space-time (\ref{canamm}).

The paper is organized as follows. In Sect. 2 we recall basic facts
concerning the twisted N-enlarged Newton-Hooke   space-times
provided in article \cite{nnh}. The third section is devoted to the short review of quantum mechanics of many particles moving in
commutative (classical) space. In Sect. 4 we construct the quantum
many-particle model defined on modified N-enlarged Newton-Hooke space-time
(\ref{canamm}).  The
final remarks are presented in the last section.

\section{Twisted N-enlarged Newton-Hooke space-times}

In this section we turn to the twisted N-enlarged Newton-Hooke space-times  equipped with two spatial directions
commuting to classical time, i.e. we consider  spaces of the form \cite{nnh}
\begin{equation}
[\;t,\hat{x}_{i}\;] =[\;\hat{x}_{1},\hat{x}_{3}\;] = [\;\hat{x}_{2},\hat{x}_{3}\;] =
0\;\;\;,\;\;\; [\;\hat{x}_{1},\hat{x}_{2}\;] =
if({t})\;\;;\;\;i=1,2,3
\;. \label{spaces}
\end{equation}
As it was already mentioned in Introduction  such a kind of quantum spaces provides the most general deformation of nonrelativistic systems. It should be noted, however,
that this type of noncommutativity  has  been  constructed explicitly  only in the case of 6-enlarged Newton-Hooke Hopf algebra, with
\begin{eqnarray}
f({t})&=&f_{\kappa_1}({t}) =
f_{\pm,\kappa_1}\left(\frac{t}{\tau}\right) = \kappa_1\,C_{\pm}^2
\left(\frac{t}{\tau}\right)\;, \nonumber\\
f({t})&=&f_{\kappa_2}({t}) =
f_{\pm,\kappa_2}\left(\frac{t}{\tau}\right) =\kappa_2\tau\, C_{\pm}
\left(\frac{t}{\tau}\right)S_{\pm} \left(\frac{t}{\tau}\right) \;,
\nonumber\\
&~~&~~~~~~~~~~~~~~~~~~~~~~~~~~~~~~~~~ \nonumber\\
&~~&~~~~~~~~~~~~~~~~~~~~~~~~~~~~~~~~~\cdot \nonumber\\
&~~&~~~~~~~~~~~~~~~~~~~~~~~~~~~~~~~~~\cdot \label{w2}\\
&~~&~~~~~~~~~~~~~~~~~~~~~~~~~~~~~~~~~\cdot \nonumber\\
&~~&~~~~~~~~~~~~~~~~~~~~~~~~~~~~~~~~~ \nonumber\\
f({t})&=&
f_{\kappa_{35}}\left(\frac{t}{\tau}\right) = 86400\kappa_{35}\,\tau^{11}
\left(\pm C_{\pm} \left(\frac{t}{\tau}\right)  \mp \frac{1}{24}\left(\frac{t}{\tau}\right)^4 - \frac{1}{2}
\left(\frac{t}{\tau}\right)^2 \mp 1\right) \,\times \nonumber\\
&~~&~~~~~~~~~~~~~~~~\times~\;\left(S_{\pm} \left(\frac{t}{\tau}\right)  \mp \frac{1}{6}\left(\frac{t}{\tau}\right)^3 - \frac{t}{\tau}\right)\;,
\nonumber\\
f({t})&=&
f_{\kappa_{36}}\left(\frac{t}{\tau}\right) =
518400\kappa_{36}\,\tau^{12}\left(\pm C_{\pm} \left(\frac{t}{\tau}\right)  \mp \frac{1}{24}\left(\frac{t}{\tau}\right)^4 - \frac{1}{2}
\left(\frac{t}{\tau}\right)^2 \mp 1\right)^2\;, \nonumber
\end{eqnarray}
and
$$C_{+/-} \left(\frac{t}{\tau}\right) = \cosh/\cos \left(\frac{t}{\tau}\right)\;\;\;{\rm and}\;\;\;
S_{+/-} \left(\frac{t}{\tau}\right) = \sinh/\sin
\left(\frac{t}{\tau}\right) \;.$$
Besides, one can easily check that in $\tau$ approaching infinity limit the above quantum spaces reproduce the canonical (\ref{noncomm}),
Lie-algebraic (\ref{noncomm1}) and quadratic (\ref{noncomm2})  type of
space-time noncommutativity, i.e. for $\tau \to \infty$ we get
\begin{eqnarray}
f_{\kappa_1}({t}) &=& \kappa_1\;,\nonumber\\
f_{\kappa_2}({t}) &=& \kappa_2\,t\;,\nonumber\\
&\cdot& \nonumber\\
&\cdot& \label{qqw2}\\
&\cdot& \nonumber\\
f_{\kappa_{35}}({t}) &=& \kappa_{35}\,t^{11}\;, \nonumber\\
f_{\kappa_{36}}({t}) &=& \kappa_{36}\,t^{12}\;. \nonumber
\end{eqnarray}
Of course, for all  parameters $\kappa_a$ $(a=1,...,36)$ running to zero the above deformations disappear.

Finally, let us notice that the spaces (\ref{spaces}) can be extended to the case of multiparticle systems as follows
\begin{equation}
[\;t,\hat{x}_{iA}\;] =[\;\hat{x}_{1A},\hat{x}_{3B}\;] = [\;\hat{x}_{2A},\hat{x}_{3B}\;] =
0\;\;\;,\;\;\; [\;\hat{x}_{1A},\hat{x}_{2B}\;] =
if({t})\delta_{AB}\;\;;\;\;i=1,2,3
\;, \label{spaces300}
\end{equation}
with $A, B = 1,2, \ldots ,M$. It should be also observed that  such an extension (blind in $A$, $B$ indecies) is compatible with canonical deformation (\ref{noncomm}).
Precisely, in $\tau$ approaching infinity limit the space (\ref{spaces300}) with function $f(t) = f_{\pm,\kappa_1}\left(\frac{t}{\tau}\right) = \kappa_1\,C_{\pm}^2
\left(\frac{t}{\tau}\right)$ passes into the well-known multiparticle canonical space-time proposed in \cite{fiore}\footnote{It should be noted that  modification of the relation (\ref{spacesfiore}) (blind in $A$, $B$ indieces as well) is in accordance with the formal arguments proposed in \cite{fiore}. Precisely, the relations (\ref{spacesfiore}) are constructed with adopt so-called braided tensor algebra procedure, dictated by structure of quantum $R$-matrix for canonical deformation \cite{qg1}, \cite{qg3}. We would like to mention, however, that in \cite{fiore} an
erroneous conclusion has been stated that based on such a twisted symmetry the noncommutative
quantum field theory (QFT) on the quantum space satisfying the relation in (\ref{noncomm}), and the usual commutative QFT are
identical. This conclusion has been reached by a misuse of the proper transformation properties of the
fields in the corresponding noncommutative space time \cite{chaia}.} (see also \cite{odynswiatowid})
\begin{equation}
[\;t,\hat{x}_{iA}\;] =[\;\hat{x}_{1A},\hat{x}_{3B}\;] = [\;\hat{x}_{2A},\hat{x}_{3B}\;] =
0\;\;\;,\;\;\; [\;\hat{x}_{1A},\hat{x}_{2B}\;] =
i\kappa_1\delta_{AB}
\;. \label{spacesfiore}
\end{equation}

\section{Quantum mechanics of many particles moving in commutative space-time  - short review}

In this section we recall basic facts concerning the  many-particle quantum mechanics defined on commutative space. First of all, we start with the following hamiltonian function
for M interacting particles
\begin{equation}
H(\bar{{p}}_1, \ldots ,\bar{{p}}_M;\bar{{r}}_1, \ldots ,\bar{{r}}_M) =
\sum_{A=1}^{M}\left(\frac{\bar{p}_A^2}{2m_A}  +V_A(\bar{r}_A)\right)
+\frac{1}{2}\sum_{A\ne B}V_{AB}(\bar{r}_A,\bar{r}_B)\;, \label{grom0}
\end{equation}
where $\bar{{p}}_{A}=[\;{{x}_{1A},{p}_{2A},{p}_{3A}}\;]$ and $\bar{{r}}_{A}=[\;{{x}_{1A},{x}_{2A},{x}_{3A}}\;]$ denote the positions and momenta operators such that
\begin{equation}
[\;x_{iA},x_{jB}\;] = 0 =[\;p_{iA},p_{jB}\;]\;\;\;,\;\;\; [\;x_{iA},p_{jB}\;]
={i\hbar}\delta_{ij}\delta_{AB}\;. \label{classpoisson}
\end{equation}
Besides, present in the above formula symbol $V_A(\bar{r}_A)$ denotes the single-particle stationary potential while $V_{AB}(\bar{r}_A,\bar{r}_B)$ describes the correlations of particles. Hence, the corresponding Schroedinger equation in so-called  position representation  looks as follows\footnote{$p_{iA} = -i\hbar \frac{\partial}{\partial x_{iA}}$.}
\begin{eqnarray}
i\frac{\partial}{\partial t} \psi(\bar{r}_1, \ldots ,\bar{r}_M,t) &=& \left[\;\sum_{A=1}^{M}\left(\frac{1}{2m_A} \Delta_A +V_A(\bar{r}_A)\right)
+\frac{1}{2}\sum_{A\ne B}V_{AB}(\bar{r}_A,\bar{r}_B)
\;\right] \times \nonumber \\
&\times& \psi(\bar{r}_1, \ldots ,\bar{r}_M,t)\;,\label{grom1}
\end{eqnarray}
 and, if one neglects
the potential functions  $V_{AB}(\bar{r}_A,\bar{r}_B)$ then, it takes the form
\begin{eqnarray}
i\frac{\partial}{\partial t} \psi(\bar{r}_1, \ldots ,\bar{r}_M,t) = \left[\;\sum_{A=1}^{M}\left(\frac{1}{2m_A} \Delta_A +V_A(\bar{r}_A)\right)
\;\right]  \psi(\bar{r}_1, \ldots ,\bar{r}_M,t)\;.\label{grom2}
\end{eqnarray}
Moreover, it is easy to see that the solution of equation (\ref{grom2}) is given by
\begin{eqnarray}
 \psi(\bar{r}_1, \ldots ,\bar{r}_M,t) = \psi_1(\bar{r}_1,t) \cdots \psi_M(\bar{r}_M,t)\;,\label{grom3}
\end{eqnarray}
with wave functions $\psi_A(\bar{r},t)$ satisfying  the standard (one-particle) differential equation
\begin{eqnarray}
i\frac{\partial}{\partial t} \psi_A(\bar{r},t) = \left(\frac{1}{2m_A} \Delta +V_A(\bar{r})\right)\psi_A(\bar{r},t)\;.\label{grom4}
\end{eqnarray}

 Usually, the potentials $V_A(\bar{r}_A)$ and $V_{AB}(\bar{r}_A,\bar{r}_B)$ remain spherically symmetric, i.e. they depend on the length of vector $\bar{r}$ and the relative positions of particles respectively
\begin{eqnarray}
V_A(\bar{r}_A) = V_A(|\bar{r}_A|)\;\;\;,\;\;\;V_{AB}(\bar{r}_A,\bar{r}_B) = V_{AB}(|\bar{r}_A - \bar{r}_B|)\;.\label{grom5}
\end{eqnarray}
Such a situation appears (for example) in the case of M electrons moving in the Coulomb field   of single nucleon with charge $Ze$ and interacting each other also by means
Coulomb potential; then, we have
\begin{eqnarray}
i\frac{\partial}{\partial t} \psi(\bar{r}_1, \ldots ,\bar{r}_M,t) &=& \left[\;\sum_{A=1}^{M}\left(\frac{1}{2m_A} \Delta_A -\frac{Ze^2}{{|\bar{r}_A|}}\right)
+\frac{1}{2}\sum_{A\ne B}\frac{e^2}{{|\bar{r}_A - \bar{r}_B|}}
\;\right] \times \nonumber \\
&\times& \psi(\bar{r}_1, \ldots ,\bar{r}_M,t)\;.\label{grom6}
\end{eqnarray}
~~~~~Finally, it should be noted that the function
\begin{eqnarray}
 \rho(\bar{r}_1, \ldots ,\bar{r}_M,t) = |\psi(\bar{r}_1, \ldots ,\bar{r}_M,t)|\;,\label{grom7}
\end{eqnarray}
can be interpreted as the density of  probability of finding first particle at point $\bar{r}_1$, second - at point $\bar{r}_2$, etc. in time-moment $t$. Besides, the average value of quantum mechanical observable ${A}$ is defined as follows
\begin{eqnarray}
<{A}> =  \int d^3{r}_1 \ldots d^3{r}_M\; \psi^*(\bar{r}_1, \ldots ,\bar{r}_M,t)\;{A}\;\psi(\bar{r}_1, \ldots ,\bar{r}_M,t)\;.\label{grom8}
\end{eqnarray}

\section{Many-body quantum mechanics for twisted N-enla-rged Newton-Hooke space-times}
~~\\
~~\\
~~\\
Let us now turn to the main aim of our investigations - to the quantum mechanical  model of many particles defined on quantum space-times (\ref{spaces300}).
In first step of our construction we extend the described in second section spaces to the whole algebra of momentum and position operators as follows
\begin{eqnarray}
&&[\;\hat{ x}_{1A},\hat{ x}_{2B}\;] = if_{\kappa_a}({t})\delta_{AB}\;\;\;,\;\;\;[\;\hat{ x}_{1A},\hat{ x}_{3B}\;] =
 [\;\hat{ x}_{2A},\hat{ x}_{3B}\;] =
 [\;\hat{ p}_{iA},\hat{ p}_{jB}\;] =0\;,\label{phasespaces1}\\
&&~~~~~~~~~~~~~~~~~[\;\hat{ x}_{iA},\hat{ p}_{jB}\;] = {i\hbar}\delta_{ij}\delta_{AB}\;\;;\;\;i,j=1,2,3\;. \label{phasespaces2}
\end{eqnarray}
One can check that relations (\ref{phasespaces1}), (\ref{phasespaces2}) satisfy the Jacobi identity and for deformation parameters
$\kappa_a$ approaching zero become classical. \\
Next, by analogy to the commutative case (see formula (\ref{grom0})) we define the following multi-particle hamiltonian operator
\begin{eqnarray}
H(\bar{\hat{p}}_1, \ldots ,\bar{\hat{p}}_M;\bar{\hat{r}}_1, \ldots ,\bar{\hat{r}}_M) =
\sum_{A=1}^{M}\left(\frac{\bar{\hat{p}}_A^2}{2m_A}  +V_A(\bar{\hat{r}}_A)\right)
+\frac{1}{2}\sum_{A\ne B}V_{AB}(\bar{\hat{r}}_A,\bar{\hat{r}}_B)\;,
\label{gromek1}
\end{eqnarray}
with $\bar{\hat{p}}_{A}=[\;{\hat{x}_{1A},\hat{p}_{2A},\hat{p}_{3A}}\;]$ and $\bar{\hat{r}}_{A}=[\;{\hat{x}_{1A},\hat{x}_{2A},\hat{x}_{3A}}\;]$. \\
In order to analyze the above system we represent the
noncommutative operators $({\hat x}_{iA}, {\hat p}_{iA})$ by classical
ones $({ x}_{iA}, { p}_{iA})$ as  (see e.g.
\cite{lodzianieosc}, \cite{lukiluk2})
\begin{eqnarray}
&~~&{\hat x}_{1A} = { x}_{1A} - \frac{1}{2\hbar}f_{\kappa_a}(t)
p_{2A}\;\;\;,\;\;\;{\hat x}_{2A} = { x}_{2A} +\frac{1}{2\hbar}f_{\kappa_a}(t)
p_{1A}\;,\\
&~~&~~~~~~~~~~~~~~~~ {\hat x}_{3A}= x_{3A} \;\;\;,\;\;\; {\hat p}_{iA}=
p_{iA}\;. \label{rep}
\end{eqnarray}
Then, the  hamiltonian (\ref{gromek1}) takes the form
\begin{eqnarray}
&~~&H(\bar{p}_1, \ldots ,\bar{p}_M;\bar{r}_1, \ldots ,\bar{r}_M,t) = \nonumber \\
&=&\sum_{A=1}^{M}\left[\;\frac{\bar{{p}}_A^2}{2m_A}  +
V_A\left(\bar{\hat{r}}_A = \left({ x}_{1A} - \frac{1}{2\hbar}f_{\kappa_a}(t)
p_{2A},{ x}_{2A} +\frac{1}{2\hbar}f_{\kappa_a}(t)
p_{1A},x_{3A}\right)\right)\right. + \nonumber \\
&~~&~~~~+\frac{1}{2}\sum_{A\ne B}V_{AB}\left(\bar{\hat{r}}_A = \left({ x}_{1A} - \frac{1}{2\hbar}f_{\kappa_a}(t)
p_{2A},{ x}_{2A} +\frac{1}{2\hbar}f_{\kappa_a}(t)
p_{1A},x_{3A}\right), \right.\label{gromek2}\\
&,&\left.\left.\bar{\hat{r}}_B = \left({ x}_{1B} - \frac{1}{2\hbar}f_{\kappa_a}(t)
p_{2B},{ x}_{2B} +\frac{1}{2\hbar}f_{\kappa_a}(t)
p_{1B},x_{3B}\right)\right)\;\right]
\;,\nonumber
\end{eqnarray}
and, consequently, the corresponding Schroedinger equation in the position representation looks as follows
\begin{eqnarray}
&~~&i\frac{\partial}{\partial t} \psi(\bar{r}_1, \ldots ,\bar{r}_M,t) = \nonumber \\
&=&\left\{\;\sum_{A=1}^{M}\left[\;\frac{1}{2m_A}\Delta_A  +
V_A\left(\bar{\hat{r}}_A = \left({ x}_{1A} + \frac{i}{2}f_{\kappa_a}(t)
{\partial_{2A}},{ x}_{2A} -\frac{i}{2}f_{\kappa_a}(t)
{\partial_{1A}},x_{3A}\right)\right)\right. \right.+ \nonumber \\
&~~&~~~~~~~+\frac{1}{2}\sum_{A\ne B}V_{AB}\left(\bar{\hat{r}}_A = \left({ x}_{1A} + \frac{i}{2}f_{\kappa_a}(t)
{\partial_{2A}},{ x}_{2A} -\frac{i}{2}f_{\kappa_a}(t)
{\partial_{1A}},x_{3A}\right), \right.\label{gromek2}\\
&,&\left.\left. \left.\bar{\hat{r}}_B = \left({ x}_{1B} + \frac{i}{2}f_{\kappa_a}(t)
{\partial_{2B}},{ x}_{2B} -\frac{i}{2}f_{\kappa_a}(t)
{\partial_{1B}},x_{3B}\right)\right)\;\right] \;\right\}\psi(\bar{r}_1, \ldots ,\bar{r}_M,t)\;.\nonumber
\end{eqnarray}
Further, we  expand the hamiltonian function (\ref{gromek2}) in Taylor series up to the terms linear in deformation parameter $\kappa_a$, i.e. to the terms linear in
function $f_{\kappa_a}(t)$; then, we have\footnote{We denote by ${\cal O}(\kappa_a)$ the higher order terms in deformation parameter $\kappa_a$.}
\begin{eqnarray}
&~~&H(\bar{p}_1, \ldots ,\bar{p}_M;\bar{r}_1, \ldots ,\bar{r}_M,t) = \nonumber \\
&=&\sum_{A=1}^{M}\left(\frac{\bar{{p}}_A^2}{2m_A}  +V_A(\bar{{r}}_A)\right)
+\frac{1}{2}\sum_{A\ne B}V_{AB}(\bar{{r}}_A,\bar{{r}}_B) +\nonumber \\
&~~&~~~~+\left[\;\sum_{A=1}^{M}\left(\left.-\frac{\partial V_A(\bar{\hat{r}}_A)}{\partial \hat{x}_{1A}}\cdot \frac{1}{2 \hbar}p_{2A}\right.\right.
+\left.\left.\frac{\partial V_A(\bar{\hat{r}}_A)}{\partial \hat{x}_{2A}}\cdot \frac{1}{2 \hbar}p_{1A}\right.\right)\right.+\nonumber\\
&+&\frac{1}{2}\sum_{A\ne B}\left.\left.\left(-\frac{\partial V_{AB}(\bar{\hat{r}}_A,\bar{\hat{r}}_B)}{\partial \hat{x}_{1A}}\cdot \frac{1}{2 \hbar}p_{2A}\right.
+ \frac{\partial V_{AB}(\bar{\hat{r}}_A,\bar{\hat{r}}_B)}{\partial \hat{x}_{2A}}\cdot\frac{1}{2 \hbar}p_{1A}\right.\right. + \label{gromek4}\\
&-&\left.\left.\left.\left.\left.\frac{\partial V_{AB}(\bar{\hat{r}}_A,\bar{\hat{r}}_B)}{\partial \hat{x}_{1B}}\cdot \frac{1}{2 \hbar}p_{2B}\right.
+ \frac{\partial V_{AB}(\bar{\hat{r}}_A,\bar{\hat{r}}_B)}{\partial \hat{x}_{2B}}\cdot \frac{1}{2 \hbar}p_{1B}\right.\right)\;\right]\right|_{f_{\kappa_a}(t)=0}\cdot f_{\kappa_a}(t) +\nonumber \\
&+& {\cal O}(\kappa_a) \;, \nonumber
\end{eqnarray}
with the corresponding  wave equation  given by
\begin{eqnarray}
&~~&i\frac{\partial}{\partial t} \psi(\bar{r}_1, \ldots ,\bar{r}_M,t) = \nonumber \\
&=&\left\{\;\sum_{A=1}^{M}\left(\frac{1}{2m_A}\Delta_A  +V_A(\bar{{r}}_A)\right)
+\frac{1}{2}\sum_{A\ne B}V_{AB}(\bar{{r}}_A,\bar{{r}}_B)\right. +\nonumber \\
&~~&~~~~+\left[\;\sum_{A=1}^{M}\left(\left.\frac{\partial V_A(\bar{\hat{r}}_A)}{\partial \hat{x}_{1A}}\cdot \frac{i}{2 }\partial_{2A}\right.\right.
\left.\left.-\frac{\partial V_A(\bar{\hat{r}}_A)}{\partial \hat{x}_{2A}}\cdot \frac{i}{2 }\partial_{1A}\right.\right)\right.+\label{gromek5}\\
&+&\frac{1}{2}\sum_{A\ne B}\left.\left.\left(\frac{\partial V_{AB}(\bar{\hat{r}}_A,\bar{\hat{r}}_B)}{\partial \hat{x}_{1A}}\cdot \frac{i}{2 }\partial_{2A}\right.
- \frac{\partial V_{AB}(\bar{\hat{r}}_A,\bar{\hat{r}}_B)}{\partial \hat{x}_{2A}}\cdot \frac{i}{2}\partial_{1A}\right.\right. + \nonumber
\end{eqnarray}
\begin{eqnarray}
&+&\left.\left.\left.\left.\left.\frac{\partial V_{AB}(\bar{\hat{r}}_A,\bar{\hat{r}}_B)}{\partial \hat{x}_{1B}}\cdot \frac{i}{2}\partial_{2B}\right.
 -\frac{\partial V_{AB}(\bar{\hat{r}}_A,\bar{\hat{r}}_B)}{\partial \hat{x}_{2B}}\cdot \frac{i}{2}\partial_{1B}\right.\right)\;\right]\right|_{f_{\kappa_a}(t)=0}\cdot f_{\kappa_a}(t) +\nonumber \\
&+& \left.{\cal O}(\kappa_a) \;\right\}\psi(\bar{r}_1, \ldots ,\bar{r}_M,t)\;. \nonumber
\end{eqnarray}
Consequently, we see that space-time noncommutativity (\ref{spaces}) generates in the hamiltonian (\ref{gromek1}) two types of additional dynamical terms. First of them arises from
the single-particle potential $V_A(\bar{\hat{r}}_A)$ while the second one corresponds to the correlations $V_{AB}(\bar{\hat{r}}_A,\bar{\hat{r}}_B)$. Of course, for deformation parameters $\kappa_a$ approaching zero all additional "potential" terms  disappear.

Let us now turn to the mentioned in pervious section the system of M particles moving "in" and interacting "by" the Coulomb potential. Then, in accordance with formulas (\ref{gromek4}) and (\ref{gromek5}) the corresponding hamiltonian function as well as the corresponding Schroedinger equation take the form
\begin{eqnarray}
H(\bar{p}_1, \ldots ,\bar{p}_M;\bar{r}_1, \ldots ,\bar{r}_M,t) &=& \sum_{A=1}^{M}\left(\frac{\bar{{p}}_A^2}{2m_A}  -\frac{Ze^2}{|{\bar{{r}}_A}|}\right)
+\frac{1}{2}\sum_{A\ne B}\frac{e^2}{{|\bar{{r}}_A - \bar{{r}}_B|}} + \nonumber\\
&~~&~~~~- \sum_{A=1}^{M} \frac{Ze^2{f_{\kappa_a}(t)}}{2\hbar{|\bar{r}_A|^3}}\cdot L_{3A} +
\label{gromek6}\\
&+&\frac{1}{2}\sum_{A\ne B}  \frac{e^2{f_{\kappa_a}(t)}}{2\hbar{|\bar{r}_A-\bar{r}_B|^3}} \cdot \left(L_{3B}+L_{3A}\right) + \nonumber\\
&-&\frac{1}{2}\sum_{A\ne B}  \frac{e^2{f_{\kappa_a}(t)}}{2\hbar{|\bar{r}_A-\bar{r}_B|^3}} \cdot \left(G_{AB}+G_{BA}\right) + {\cal O}(\kappa_a)\;,\nonumber
\end{eqnarray}
and
\begin{eqnarray}
i\frac{\partial}{\partial t} \psi(\bar{r}_1, \ldots ,\bar{r}_M,t) &=& \left[\;\sum_{A=1}^{M}\left(\frac{1}{2m_A}\Delta_A  -\frac{Ze^2}{|{\bar{{r}}_A}|}\right)
+\frac{1}{2}\sum_{A\ne B}\frac{e^2}{{|\bar{{r}}_A - \bar{{r}}_B|}} + \right.\nonumber\\
&~~&~~~~ -\sum_{A=1}^{M} \frac{Ze^2{f_{\kappa_a}(t)}}{2\hbar{|\bar{r}_A|^3}}\cdot L_{3A} +
\label{gromek7}\\
&+&\left.\frac{1}{2}\sum_{A\ne B} \frac{e^2{f_{\kappa_a}(t)}}{2\hbar{|\bar{r}_A-\bar{r}_B|^3}}\cdot \left(L_{3B}+L_{3A}\right) +\;\right.  \nonumber \\
&-&\left.\frac{1}{2}\sum_{A\ne B} \frac{e^2{f_{\kappa_a}(t)}}{2\hbar{|\bar{r}_A-\bar{r}_B|^3}}\cdot \left(G_{AB}+G_{BA}\right) + {\cal O}(\kappa_a)\;\right]\times \nonumber \\
&\times& \psi(\bar{r}_1, \ldots ,\bar{r}_M,t)\;.\nonumber
\end{eqnarray}
respectively, with $L_{3A} =x_{1A}p_{2A} - x_{2A}p_{1A}$ and $G_{AB} = x_{1B}p_{2A} - x_{2B}p_{1A}$. Particulary, in the case of single particle, for canonical deformation $f_{\kappa_a}(t) = \kappa_a$  we reproduce
the noncommutative model of hydrogen atom proposed in \cite{qm1} and \cite{qm2}
\begin{eqnarray}
H(\bar{p},\bar{x}) &=& \frac{\bar{p}^2}{2m} -\frac{Ze^2}{{|\bar{r}|}} - \frac{Ze^2{\kappa_a}}{2\hbar{|\bar{r}|^3}}\cdot L_3 + \mathcal{O}
(\kappa_a) \;,\label{gromek8}
\end{eqnarray}
while for more complicated (time-dependent) functions $f_{\kappa_a}(t)$, we get the one-particle system described by
\begin{eqnarray}
H(\bar{p},\bar{x},t) &=& \frac{\bar{p}^2}{2m} -\frac{Ze^2}{{|\bar{r}|}} - \frac{Ze^2f_{\kappa_a}(t)}{2\hbar{|\bar{r}|^3}}\cdot L_3 + \mathcal{O}
(\kappa_a) \;. \label{gromek9}
\end{eqnarray}
It is well-known, that the solution of the  corresponding (associated with (\ref{gromek9})) Schroedinger equation  can be found  with use of time-dependent perturbation theory \cite{24}. It looks as follows
\begin{eqnarray}
 {\psi}(\bar{r},t) = \sum_{n=0}^{\infty}\sum_{l=0}^{n-1}\sum_{m=-l}^{l} c_{nlm}(t){\rm e}^{iE_n(t-t_0)}\psi_{nlm}(\bar{x})\;, \label{gromek10}
\end{eqnarray}
where symbols $E_n$ and $\psi_{nlm}$ denote eigenvalues and eigenfunctions for hydrogen atom, while coefficients $c_{nlm}(t)$ are defined as the solutions of the following differential equations
\begin{eqnarray}
\frac{dc_{nlm}(t)}{dt} &=& -\frac{1}{i\hbar} \sum_{n'=0}^{\infty}\sum_{l'=0}^{n-1}\sum_{m'=-l}^{l} \left(\psi_{nlm}(\bar{r}),\frac{Ze^2f_{\kappa_a}(t)}{2\hbar{|\bar{r}|^3}}\cdot L_3\psi_{n'l'm'}(\bar{r})\right)c_{n'l'm'}(t_0)\;\cdot \cr
&\cdot&{\rm e}^{i\omega_{nn'}(t-t_0)}\;\;\;;\;\;\;\omega_{nn'} \;=\; \frac{1}{\hbar}(E_n-E_{n'})\;.\label{gromek11}
\end{eqnarray}
Hence, in accordance with prescription (\ref{grom3}),  the solution of  multiparticle wave equation (\ref{gromek7}) with neglected correlation potential $V_{AB}(|\bar{r}_A-\bar{r}_B|)$ and vanishing $\mathcal{O}
(\kappa_a)$-terms takes the form
\begin{eqnarray}
 \psi(\bar{r}_1, \ldots ,\bar{r}_M,t) = {\psi}_1(\bar{r}_1,t) \cdots {\psi}_M(\bar{r}_M,t)\;,\label{gromek3}
\end{eqnarray}
with one-particle functions ${\psi}_A(\bar{r}_A,t)$ given by (\ref{gromek10}).

Finally, it should be noted that the average values of energy operators (\ref{gromek2}), (\ref{gromek4}) and (\ref{gromek6}) can be found with use of the formula (\ref{grom8}).

\section{Final remarks}

In this article we construct the quantum model of M
nonrelativistic particles moving in noncommutative space-time
(\ref{spaces300}). The corresponding Schroedinger equation for arbitrary stationary
potential is provided and, in
particular, there is analyzed the distinguished example of such
system - the set of M particles moving "in" and interacting "by" the Coulomb potential. It
should be noted, however, that by analogy to the investigations performed in article \cite{qm1}, one can
ask about more physical features (such as for example the energy spectrum or the Lamb shift) of the model defined by Hamiltonian (\ref{gromek6}).
Besides,  it should be added, that the presented  considerations give a starting
point for the construction of Dirac quantum mechanics for
many particles defined on the relativistic counterpart of modified space-time (\ref{canamm}).
 The studies in these directions  already started and
are in progress.

\section*{Acknowledgments}
The author would like to thank J. Lukierski
for valuable discussions. This paper has been financially  supported  by Polish
NCN grant No 2011/01/B/ST2/03354.

\end{document}